# Constrained Variational Autoencoder for improving EEG based Speech Recognition Systems

*Gautam Krishna, Co Tran, Mason Carnahan, Ahmed H Tewfik*

Brain Machine Interface Lab, The University of Texas at Austin

## Abstract

In this paper we introduce a recurrent neural network (RNN) based variational autoencoder (VAE) model with a new constrained loss function that can generate more meaningful electroencephalography (EEG) features from raw EEG features to improve the performance of EEG based speech recognition systems. We demonstrate that both continuous and isolated speech recognition systems trained and tested using EEG features generated from raw EEG features using our VAE model results in improved performance and we demonstrate our results for a limited English vocabulary consisting of 30 unique sentences for continuous speech recognition and for an English vocabulary consisting of 2 unique sentences for isolated speech recognition. We compare our method with another recently introduced method described by authors in [1] to improve the performance of EEG based continuous speech recognition systems and we demonstrate that our method outperforms their method as vocabulary size increases when trained and tested using the same data set.

Even though we demonstrate results only for automatic speech recognition (ASR) experiments in this paper, the proposed VAE model with constrained loss function can be extended to a variety of other EEG based brain computer interface (BCI) applications.

## 1. Introduction

The emergence of virtual personal assistants like Amazon Alexa, Apple Siri, Windows Cortana, Samsung Bixby etc has improved the interactions between user and smart phones, personal computers etc. Most of the virtual personal assistants support voice recognition which allows the users to interact with them hands free. However current voice recognition systems used in virtual personal assistants are trained to recognize only acoustic features and their performance degrades in presence of background noise as well as when the input speech is distorted or broken. This limits technology accessibility for smart phone users with speaking disabilities or for people who are not able to produce speech. Recently researchers have started exploring the possibility of using deep learning models to perform isolated and continuous speech recognition using non invasive electroencephalography (EEG) neural signals as demonstrated by the works explained in [2, 3, 4]. The EEG is a non invasive way of measuring electrical activity of human brain. The EEG sensors are placed on the scalp of the subject to obtain the recordings. The EEG signals offer very high temporal resolution and the technique is completely non invasive making it extremely safe and easily deployable.

In [3] authors demonstrated isolated speech recognition using EEG signals on a limited English vocabulary of four words and five vowels with high accuracy and they also demonstrated that EEG features can be used to improve the performance of isolated speech recognition systems operating in presence of background noise. In [2] authors demonstrated continuous speech recognition using EEG features for a limited English vocabulary consisting of 30 unique sentences. Similarly in [4] authors demonstrated continuous speech recognition using various EEG features sets for a limited vocabulary of 9 unique English sentences. The results demonstrated by authors in [2, 4] show that continuous speech recognition using EEG features is more challenging than isolated speech recognition using EEG [3].

Recently in [1] authors introduced various techniques to improve the performance of EEG based continuous speech recognition systems. They demonstrated that by initializing the weights of the encoder layers in the automatic speech recognition (ASR) model with weights from a regression model trained to predict concatenation of acoustic and articulatory features from EEG features will help in improving the performance of the ASR system for performing recognition using EEG features.

In [5, 6] authors introduced the concept of variational autoencoder (VAE). The VAE is related to a normal autoencoder [7] but with more constraints or control on latent representation. A normal autoencoder learns the compressed representation of data automatically by first compressing the input and decompressing it back to match the original input. The compression part is done by the encoder model and the decompression part is done by the decoder model in the autoencoder model. A VAE operates in the same way like the autoencoder but a VAE models the latent variables as isotropic gaussian priors thus allowing each dimension in the latent representation to be as independent as possible [5]. In this we paper we demonstrate that by making use of this independence property of the latent representation modeling in VAE and by adding a new term to the VAE loss function, the model can be used to generate more meaningful EEG features from raw EEG features to improve the performance of EEG based speech recognition systems. We compare our method with the method described by authors in [1] to improve the performance of EEG based continuous speech recognition systems and we demonstrate that our method outperforms their method for larger test set vocabulary sizes, when trained and tested using the same data set and our proposed method doesn't need additional features like acoustic features or articulatory features which are needed to implement the method described by authors in [1].

## 2. Variational Autoencoder (VAE) model with constrained loss function

The variational autoencoder (VAE) like a normal autoencoder learns the compressed representation of data automatically by first compressing the input and decompressing it back to match the original input but the latent representation in VAE is modeled as isotropic gaussian priors thus allowing each dimension in the latent representation to be as independent as possible [5]. The overview of our idea is described in Figure 1. The basic idea is to modify the VAE model in such a way that it can take raw recorded experimental EEG features as input, denoises the EEG features and generates EEG features which are the best representations of acoustic features, since we are only interested in EEG features which are helpful in improving the performance of ASR systems. The Figure 2 explains the architecture of our model in detail. As seen from Figure 1, the encoder in the VAE model takes EEG features recorded in parallel with speech as input, transforms it into latent representation and the decoder model reconstructs the EEG features from the latent space. We used latent space of dimension 5. The latent space dimension value is a hyper parameter and it was chosen to be set to the value of 5 based on hyper parameter tuning experiments. Based on the property of VAE, each dimension in the latent space tries to be as independent as possible. Even though the input raw EEG features were recorded in parallel with speech, during speaking process, the subject's brain is not only processing speech production but in parallel it is processing other activities like emotions, thoughts etc, hence the EEG neural recordings reflects a mixture of brain activity responsible for speech production, emotions, thoughts etc. In order to design a robust reliable EEG based speech prosthetic we should be able to separate out EEG activity responsible only for speech production from the rest, hence any one dimension output from the latent space ( in our work we used the last node output from latent space as seen from Figure 1) is passed to a ASR model in every epoch during the training of VAE model. So now the model will have a net loss consisting of the VAE default loss plus the ASR model loss. The intuition here is as the VAE model and ASR model are trained simultaneously, the four other dimensions in the latent space of the VAE will learn the representations of neural activities responsible for non speaking related activities and the fifth node or dimension (that is connected to the ASR model) in the latent space will learn to generate EEG features which are the closest representation of acoustic features or in other words the fifth dimension in latent space will produce EEG features or neural activities responsible only for speaking task as the both ASR and VAE models are trained simultaneously until the combined net loss is showing convergence.

As seen from Figure 2, the encoder of our VAE model is a single layer long short-term memory (LSTM) [8] with 128 hidden units which takes raw EEG features of dimension 30 as input. The last time step output of encoder LSTM is passed to dense layers consisting of hidden units same as number of time steps of EEG. The number of time steps of EEG is computed as the product of the sampling frequency of EEG features and sequence length. The dense layer outputs are then repeated for 5 times, where 5 corresponds to the latent space dimension to form the $Z_{mean}$ (mean) and $Z_{LogSigma}$ (variance) vectors. Using $Z_{mean}$ and $Z_{LogSigma}$ a point from the latent space is sampled [6]. The output of the sampling layer is of the shape [batch size, 5, time steps] and it is then reshaped to [batch size, 5, time steps, 1]. Then the fifth dimension value from latent space (from the previous reshaped tensor) of the form [batch size, time steps, 1] is fed into the ASR classifier model described in Figure 3 during every training epoch. As seen from Figure 3 the ASR classifier model consists of two layer of gated recurrent unit (GRU) [9] with 128, 64 hidden units respectively with dropout [10] regularization followed by a linear dense layer consisting of 64 hidden units followed by a final dense layer with 2 hidden units and softmax activation function to get prediction probabilities. The loss function of the ASR classifier model was categorical cross entropy.

The sampled point is decoded using a two layer LSTMs with 128 and 30 hidden units respectively and the reconstruction error or mean squared error (MSE) between the decoded EEG features and input EEG features is computed as shown in Figure 2. The VAE also have an additional KL Divergence loss computed between the returned distribution and a standard Gaussian to make distributions returned by the encoder LSTM close to a standard normal distribution [5, 6]. The model's net training loss convergence is shown in Figure 4. Both the VAE and ASR classifier models were trained simultaneously for 100 epochs with a batch size of one using rmsprop as the optimizer. During test time the LSTM encoder in the trained VAE model takes EEG features of dimension 30 as input and we take output from the fifth dimension node in the VAE latent space which outputs EEG features of dimension one.

## 3. ASR models used for performing experiments

We performed both isolated and continuous speech recognition using the raw EEG features of dimension 30 (baseline) and also using the EEG features of dimension one generated using the fifth dimension node in the latent space of the VAE model described before. For performing isolated speech recognition we used an ASR classifier model similar to the one explained in Figure 3 but instead of GRU (128) we used temporal convolutional network (TCN) [11] layer with 128 filters and instead of the GRU (64) layer a GRU (32) layer was used and we skipped the Dense (64) units layer. The classifier model was trained for 200 epochs with batch size 200, categorical cross entropy as loss function and using adam [12] as the optimizer.

For performing continuous speech recognition experiments we used the connectionist temporal classification (CTC) [13, 14] model described in Figure 1 in [1] with the exact same hyper parameters and training parameters used by authors in [1] but the encoder layers in the CTC model were initialized with random weights [2, 4]. An external language model was used during inference time like the ones used by authors in [1].

## 4. Data Sets used for performing experiments

For performing continuous speech recognition experiments using EEG we used Data set B used by authors in [1]. First we perform continuous speech recognition experiments using EEG features of dimension 30 from Data set B [1] and then we pass the EEG features of dimension 30 to our model described in Figure 2 and get the output from fifth dimension node of the latent space to get EEG features of dimension one. Then experiments are performed using those EEG features of dimension one.

For performing isolated speech recognition experiments we used the combined EEG data for first two unique sentences from Data set A and B used by authors in [2], consisting of a total of 108 EEG recording examples. The same data set was used to train our VAE model described in Figure 2. Since there were only two unique sentences, hence the ASR classifier model's final dense layer had two hidden units with softmax activation function. We considered EEG samples for only two unique sentences since we were interested in faster training of the simulta-neous VAE and ASR classifier models. More details of the data set, EEG experiment design, EEG recording hardware etc are covered in [2, 1].

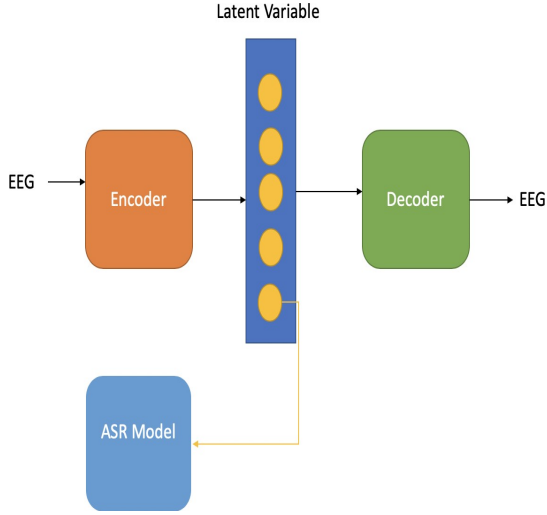

Figure 1: *Overview of our proposed VAE model*

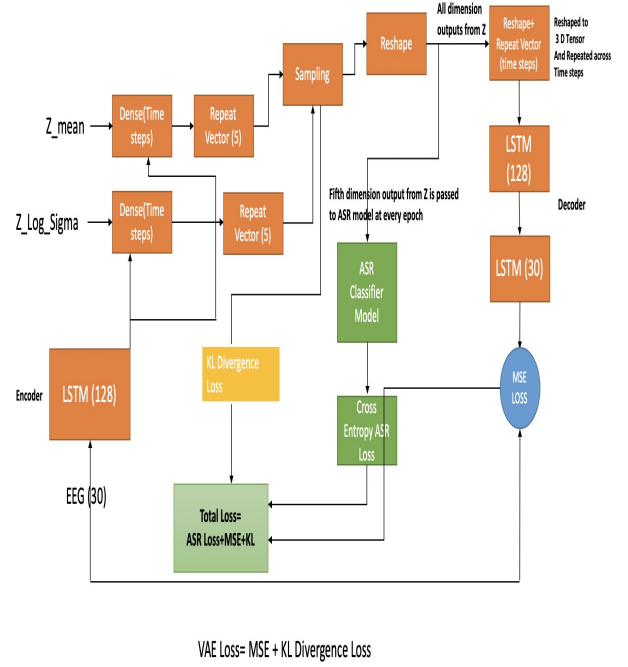

Figure 2: *Architecture of our proposed VAE model*

For each data set we used 80% of the data as training set and rest as test set. The train-test split was done randomly. There was no overlap between training and testing set. The way we splitted data for performing continuous speech recognition experiments in this work was exactly similar to the method used by authors in [1].

## 5. EEG feature extraction details

We followed the same EEG preprocessing methods used by authors in [3, 2] for extracting raw EEG features. The EEG signals were sampled at 1000Hz and a fourth order IIR band pass filter with cut off frequencies 0.1Hz and 70Hz was applied. A notch filter with cut off frequency 60 Hz was used to remove the power line noise. The EEGlab's [15] Independent component analysis (ICA) toolbox was used to remove other biological signal artifacts like electrocardiography (ECG), electromyography (EMG), electrooculography (EOG) etc from the EEG signals. We extracted five statistical features for EEG, namely root mean square, zero crossing rate, moving window average, kurtosis and power spectral entropy [3, 2]. So in total we extracted 31(channels) X 5 or 155 features for EEG signals. The EEG features were extracted at a sampling frequency of 100Hz for each EEG channel.

## 6. EEG Feature Dimension Reduction Algorithm Details

After extracting EEG features as explained in the previous section, we used Kernel Principle Component Analysis (KPCA) [16] to perform initial denoising of the EEG feature space as explained by authors in [2, 3]. We reduced the 155 EEG features to a dimension of 30 by applying KPCA for both the data sets. We plotted cumulative explained variance versus number of components to identify the right feature dimension. We used KPCA with polynomial kernel of degree 3 [3, 2]. We used these EEG features of dimension 30 as EEG features for calculating baseline results for both isolated and continuous speech recognition experiments and these 30 EEG dimensional EEG features are passed to the model described in Figure 2 to get EEG features of dimension one.

## 7. Results

For isolated speech recognition experiments during test time we used classification accuracy as the performance metric and for continuous speech recognition experiments during test time we used word error rate (WER) as performance metric. The classification accuracy during test time is defined as the ratio of number of correct predictions given by the model to total number of predictions given by model on test set. The results obtained for isolated speech recognition during test time are described in Table 1. As seen from Table 1, training and testing isolated speech recognition classifier model with EEG features generated using the fifth dimension in the latent space of our VAE model resulted in **4.55 %** performance improvement compared to the baseline where the model was trained and tested using 30 dimensional EEG features. The isolated speech recognition results demonstrated by authors in [3] had higher accuracy as they had more examples per each label compared to our set up and in their case labels were vowels and words but in our case label is complete sentence.

Table 2 shows the test time results obtained for continuous speech recognition experiments. For baseline results we use 30 dimensional EEG features with CTC encoder with random weights, we then compare results obtained using our proposed method in this paper with the results obtained by authors in [1]. We specifically compare our results with the results explained in Table 2 in reference [1]. As seen from Table 2 continuous speech recognition using EEG features of dimension one generated using our model described in Figure 2 always resulted in superior performance compared to baseline and demonstrated superior performance or lower WER compared to the method introduced by authors in [1] for larger test set corpus sizes.

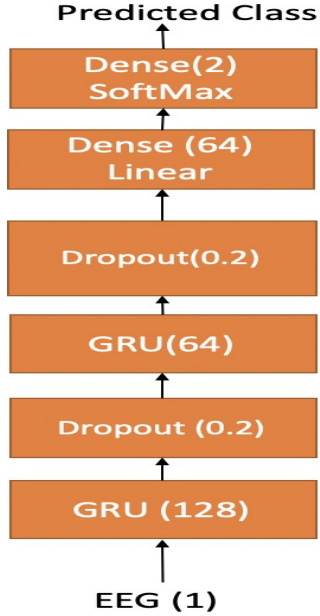

Figure 3: *Architecture of the ASR classifier model used in our proposed VAE model*

| EEG DIM 30 BASELINE (% Test Accuracy) | EEG DIM 1 (% Test Accuracy) |
|---|---|
| 50 | 54.55 |

Table 1: *Test time results for **isolated speech recognition***

| Total Number of Sentences | WER (%) EEG DIM 30 BASELINE | WER (%) EEG DIM 30 REF [1] TECHNIQUE | WER (%) EEG DIM 1 PROPOSED TECHNIQUE |
|---|---|---|---|
| 30 | 82.63 | 74.36 | 75.47 |
| 60 | 84.30 | 74.45 | 77.57 |
| 90 | 82.67 | 77.76 | 79.85 |
| 120 | 88.94 | 79.68 | **74.48** |
| 150 | 90.39 | 81.97 | **78.15** |
| 180 | 85.39 | 84.9 | **84.22** |

Table 2: *Test time results for **continuous speech recognition***

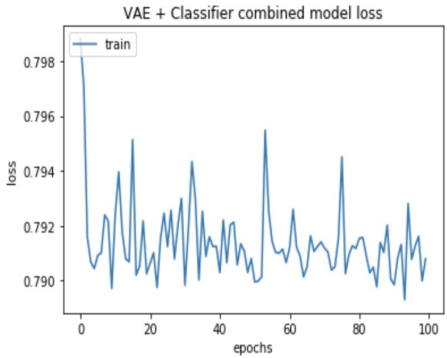

Figure 4: *Training loss*

Results from Tables 1 and 2 summarizes that our proposed method can be used to generate EEG features to improve the performance of isolated and continuous EEG based speech recognition systems. Our proposed method doesn't depend on additional features like acoustic or articulatory features like the method used by authors in [1].

## 8. Conclusion and Future work

In this we paper we demonstrate that by making use of the independent and identically distributed (IID) property of the latent representation modeling in VAE and by adding a ASR loss term to the VAE loss function, the model can be used to generate more meaningful EEG features from raw EEG features to improve the performance of EEG based speech recognition systems. We compare our method with the method described by authors in [1] to improve the performance of EEG based continuous speech recognition systems and we demonstrate that our method outperforms their method for larger test set vocabulary sizes, when trained and tested using the same data set and our proposed method doesn't need additional features like acoustic features or articulatory features which are needed to implement the method described by authors in [1].

For future work we would like to improve the current results by replacing the ASR classifier model in the VAE setup with continuous speech recognition models like CTC or Attention model but that will require larger training data set with more number of EEG examples and data from larger number of subjects. We would also like to combine our proposed method with the method introduced by authors in [1] to see if that helps in establishing a new baseline for state-of-the-art continuous EEG based speech recognition.

Finally, even though we demonstrate results only for automatic speech recognition (ASR) experiments in this paper, the proposed VAE model with constrained loss function can be extended to a variety of other EEG based brain computer interface (BCI) applications by using other application specific loss functions instead of ASR loss in the VAE set-up.

## 9. Acknowledgement

We would like to thank Kerry Loader and Rezwanul Kabir from Dell, Austin, TX for donating us the GPU to train the models used in this work. The first author would like to thank Shanshan Wu from Google Research (initially at UT Austin) for recommending VAE approach over normal AE to solve this problem.

# 10. References

[1] G. Krishna, C. Tran, M. Carnahan, Y. Han, and A. H. Tewfik, "Improving eeg based continuous speech recognition," *arXiv preprint arXiv:1911.11610*, 2019.

[2] G. Krishna, C. Tran, M. Carnahan, and A. Tewfik, "Advancing speech recognition with no speech or with noisy speech," in *2019 27th European Signal Processing Conference (EUSIPCO)*. IEEE, 2019.

[3] G. Krishna, C. Tran, J. Yu, and A. Tewfik, "Speech recognition with no speech or with noisy speech," in *Acoustics, Speech and Signal Processing (ICASSP), 2019 IEEE International Conference on*. IEEE, 2019.

[4] G. Krishna, Y. Han, C. Tran, M. Carnahan, and A. H. Tewfik, "State-of-the-art speech recognition using eeg and towards decoding of speech spectrum from eeg," *arXiv preprint arXiv:1908.05743*, 2019.

[5] D. P. Kingma and M. Welling, "Auto-encoding variational bayes," *arXiv preprint arXiv:1312.6114*, 2013.

[6] D. P. Kingma, M. Welling *et al.*, "An introduction to variational autoencoders," *Foundations and Trends® in Machine Learning*, vol. 12, no. 4, pp. 307–392, 2019.

[7] D. E. Rumelhart, G. E. Hinton, and R. J. Williams, "Learning internal representations by error propagation," California Univ San Diego La Jolla Inst for Cognitive Science, Tech. Rep., 1985.

[8] S. Hochreiter and J. Schmidhuber, "Long short-term memory," *Neural computation*, vol. 9, no. 8, pp. 1735–1780, 1997.

[9] J. Chung, C. Gulcehre, K. Cho, and Y. Bengio, "Empirical evaluation of gated recurrent neural networks on sequence modeling," *arXiv preprint arXiv:1412.3555*, 2014.

[10] N. Srivastava, G. Hinton, A. Krizhevsky, I. Sutskever, and R. Salakhutdinov, "Dropout: a simple way to prevent neural networks from overfitting," *The journal of machine learning research*, vol. 15, no. 1, pp. 1929–1958, 2014.

[11] S. Bai, J. Z. Kolter, and V. Koltun, "An empirical evaluation of generic convolutional and recurrent networks for sequence modeling," *arXiv preprint arXiv:1803.01271*, 2018.

[12] D. P. Kingma and J. Ba, "Adam: A method for stochastic optimization," *arXiv preprint arXiv:1412.6980*, 2014.

[13] A. Graves, S. Fernández, F. Gomez, and J. Schmidhuber, "Connectionist temporal classification: labelling unsegmented sequence data with recurrent neural networks," in *Proceedings of the 23rd international conference on Machine learning*. ACM, 2006, pp. 369–376.

[14] A. Graves and N. Jaitly, "Towards end-to-end speech recognition with recurrent neural networks," in *International Conference on Machine Learning*, 2014, pp. 1764–1772.

[15] A. Delorme and S. Makeig, "Eeglab: an open source toolbox for analysis of single-trial eeg dynamics including independent component analysis," *Journal of neuroscience methods*, vol. 134, no. 1, pp. 9–21, 2004.

[16] S. Mika, B. Schölkopf, A. J. Smola, K.-R. Müller, M. Scholz, and G. Rätsch, "Kernel pca and de-noising in feature spaces," in *Advances in neural information processing systems*, 1999, pp. 536–542.